2018-09-06

# Optical frequency comb photoacoustic spectroscopy

Ibrahim Sadiek,[a] Tommi Mikkonen,[b] Markku Vainio,[b,c] Juha Toivonen,[b] Aleksandra Foltynowicz*[a]

We report the first photoacoustic detection scheme using an optical frequency comb — the optical frequency comb photoacoustic spectroscopy (OFC-PAS). OFC-PAS combines the broad spectral coverage and the high resolution of OFCs with the small sample volume of cantilever-enhanced PA detection. In OFC-PAS, a Fourier transform spectrometer (FTS) is used to modulate the intensity of the exciting comb source at a frequency determined by its scanning speed. One of the FTS outputs is directed to the PA cell and the other is measured simultaneously with a photodiode and used to normalize the PA signal. The cantilever-enhanced PA detector operates in a non-resonant mode, enabling detection of broadband frequency response. The broadband and the high-resolution capabilities of OFC-PAS are demonstrated by measuring the rovibrational spectra of the fundamental C-H stretch band of $CH_4$, with no instrumental line shape distortions, at total pressures of 1000 mbar, 650 mbar, and 400 mbar. In this first demonstration, a spectral resolution two orders of magnitude higher than previously reported with broadband PAS is obtained, limited by the pressure broadening. A limit of detection of 0.8 ppm of methane in $N_2$ is accomplished in a single interferogram measurement (200 s measurement time, 1000 MHz spectral resolution, 1000 mbar total pressure) for an exciting power spectral density of 42 $\mu$W/cm$^{-1}$. A normalized noise equivalent absorption of $8 \times 10^{-10}$ W cm$^{-1}$ Hz$^{-1/2}$ is obtained, which is only a factor of three higher than the best reported with PAS based on continuous wave lasers. A wide dynamic range of up to four orders of magnitude and a very good linearity (limited by the Beer-Lambert law) over two orders of magnitude are realized. OFC-PAS extends the capability of optical sensors for multispecies trace gas analysis in small sample volume with high resolution and selectivity.

## 1 Introduction

Photoacoustic spectroscopy (PAS) is a routine technique in many vital applications including trace gas analysis,[1] and characterization of industrial products.[2] The small sample volume and zero-background are unique advantages over conventional absorption-based techniques (e.g., direct and cavity-enhanced absorption spectroscopy). Since the discovery of the photoacoustic effect by Bell in 1880,[3] and the seminal work of Kerr and Atwood on the first PA detection of infrared absorption in gases,[4] the performance of light sources, modulators, and detectors used for PAS has been significantly improved. High-power continuous wave (cw) laser sources in combination with different PAS detection schemes allow monitoring of trace gases in real time,[1,5] and detection limits down to parts-per-trillion.[6–8] However, cw lasers can be tuned only over a narrow spectral bandwidth, which may limit their usage to single species detection. Attention has also been paid to the development of new sensing elements, replacing the conventional condenser and electret microphones used to detect the optically-induced acoustic wave.[9–15] In quartz-enhanced PAS,[10–12] a quartz tuning fork with resonance frequency ~32 kHz is used as an extremely sensitive acoustic detector because of its high quality factor (Q-factor > $10^4$). However, the commitment to resonance introduces restrictions on light sources because of the need for accurate control of modulation frequency. In contrast, cantilever-enhanced detectors[13–16] can be operated in a non-resonant mode, enabling detection of a wide range of audio frequencies,[13] and thus broadband detection schemes. In addition, they offer better mechanical stability and larger movement range than the flexible membranes in capacitive microphone detectors.[13]

Broadband detection of PA signals has so far been performed mostly with incoherent infrared radiators (e.g., lamps, blackbodies, etc.) modulated by conventional Fourier transform infrared (FT-IR) spectrometers.[15,17] When broadband radiation is passing through a scanning Michelson interferometer, each wavenumber component, $\nu$, is modulated at its characteristic Fourier frequency, $f = V \times \nu$, where $V$ is the optical path difference (OPD) scan velocity. Absorption at each $\nu$ is manifested as an acoustic wave at the corresponding modulation frequency that can be measured with the microphone. First broadband FT-IR-PAS instruments used electret microphones and were not widely applied because of the low spectral irradiance of broadband IR sources and the low sensitivity of the microphones.[12,13,15] The implementation of cantilever-enhanced detectors significantly increased the detection sensitivity and FT-IR-PAS setups with detection limits of 1.5 ppm (3$\sigma$, 100 s, spectral resolution 4 cm$^{-1}$),[18] and 3 ppm (2$\sigma$, 168 s, spectral resolution 8 cm$^{-1}$)[15] have been reported for methane. In a very recent demonstration, Mikkonen et al.[19] used a supercontinuum (SC) light source for broadband cantilever-enhanced PA detection and reported a limit of detection for methane of 1.4 ppm (3$\sigma$, 50 s, spectral resolution 4 cm$^{-1}$). Although the brightness of SC sources exceeds that of thermal emitters by orders of magnitude, the noise level is also increased, which is the reason for the relatively small improvement of the detection limit.

The spectral resolution of all broadband PAS demonstrations has so far been limited by the nominal resolution of the FT-IR spectrometer, given by the inverse of the maximum delay range.[20] Conventional FT-IR spectrometers provide a resolution of 0.1 $cm^{-1}$ (i.e., 3 GHz), and higher resolutions, up to 0.002 $cm^{-1}$, have been achieved with large instruments (OPD = 4.5 m) and long measurement times (30 min).[21] In addition, care has to be taken to select a proper apodization function in order to minimize the instrumental line shape (ILS) distortions. Using an optical frequency comb (OFC) as a light source for Fourier transform spectrometers (FTS), in the so-called comb-based FTS, allows much faster acquisition (of the order of seconds) of spectra with high signal-to-noise ratio (*S*/*N*),[22] and resolution down to kHz range with no ILS contribution.[23]

[a] Department of Physics, Umeå University, 901 87 Umeå, Sweden.
E-mail: aleksandra.foltynowicz@umu.se.
[b] Laboratory of Photonics, Tampere University of Technology, Tampere, Finland.
[c] Department of Chemistry, University of Helsinki, Finland.

Here, we report the first demonstration of photoacoustic spectroscopy using an optical frequency comb. We call the technique: optical frequency comb photoacoustic spectroscopy (OFC-PAS). OFC-PAS combines the wide spectral coverage and high resolution of comb-based FTS with the small sample volume and wide dynamic range of photoacoustic detection. In this first demonstration, we measured high resolution PA spectra of the fundamental C-H stretch band of methane, $CH_4$, in nitrogen, $N_2$, at different pressures, in good agreement with simulations based on the HITRAN database.[24] Spectral resolution up to 400 MHz, limited by the pressure broadening and 75 times better than the best reported using a SC source (30 GHz),[19] has been accomplished. These capabilities open up for new insights for high resolution multicomponent trace gas analysis in small sample volumes.

## 2 Experimental

A schematic of the OFC-PAS setup is shown in Fig. 1. The mid-IR optical frequency comb is based on a doubly-resonant optical parametric oscillator (DROPO) with an orientation-patterned GaAs (OP-GaAs) crystal. The DROPO is pumped by a Tm:fibre femtosecond laser (IMRA America) around 1.95 μm with a repetition rate of 125 MHz and up to 2.5 W output power. The DROPO can be operated in both degenerate and non-degenerate modes. In the non-degenerate mode, the DROPO output consists of a signal comb with ~250 nm bandwidth and centre frequency tunable between 3.1 – 3.6 μm, and an idler comb with ~350 nm bandwidth tunable between 4.6 – 5.4 μm. Here we use the signal comb light around 3.3 μm (3000 $cm^{-1}$) for excitation of the fundamental C-H stretch band of $CH_4$. A detailed description of the DROPO operation is given in ref.[25] Compared to that work, the coupling efficiency of the pump laser power into the DROPO cavity has been optimized by better choice of mode-matching lenses, providing up to 60 mW of signal and idler power compared to 48 mW obtained before. Moreover, the bandwidth of the dither lock used for stabilizing the DROPO cavity length has been increased to increase the output power and decrease the intensity noise. This has been achieved by mounting one of the DROPO cavity mirrors on a fast (70 kHz bandwidth) piezoelectric transducer (PZT). The fast PZT is used to dither the cavity length at 77 kHz, while both the fast and the slow (12 kHz bandwidth) PZTs are used for locking. The synchronous demodulation of the output of a PbSe detector, placed after the input coupler, yields an error signal that is fed back to the PZTs via two servo controllers.

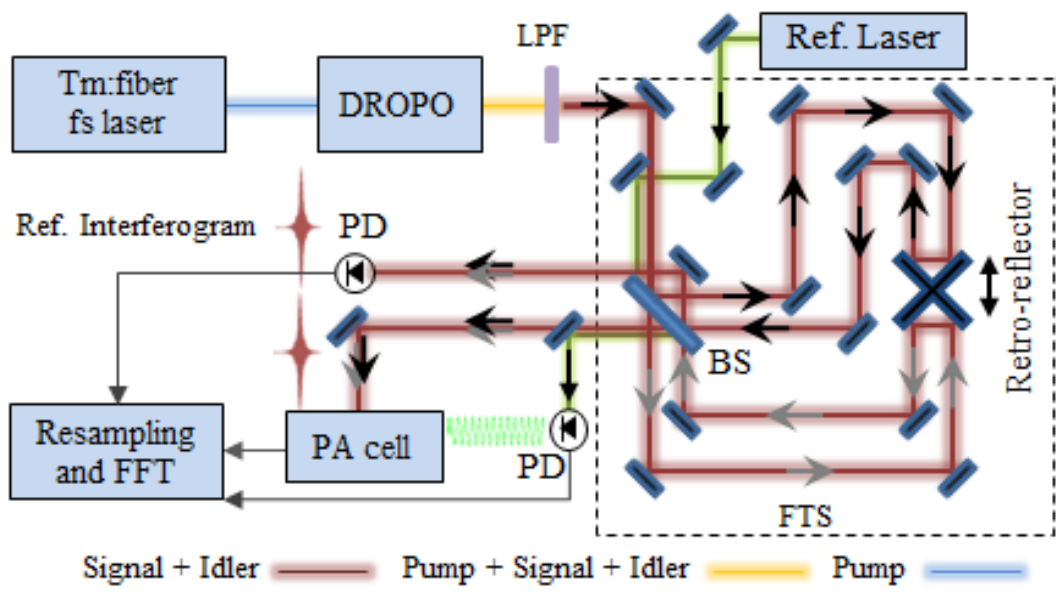

Fig. 1 Schematic of the experimental setup: DROPO – doubly resonant optical parametric oscillator; LPF – long-pass filter; BS – beam splitter; PD – photodetector; PA – photoacoustic; FTS – Fourier transform spectrometer.

The output of the DROPO, after long-pass filter in order to block the remainder of the pump power, is coupled into a home-built fast-scanning FTS. A general schematic of the tilt-compensated interferometer is shown in Fig. 1 (dashed box). A stable cw reference diode laser ($\lambda_{ref}$ = 1563 nm) is also coupled to the FTS and is used for the calibration of the OPD. The two retro-reflectors, mounted back-to-back, provide easier alignment and larger OPD range compared to conventional interferometers with one moving mirror. The maximum OPD range is 2.8 m, corresponding to a nominal resolution of 0.0037 $cm^{-1}$, or 110 MHz. The OPD is scanned at $V$ = 0.16 cm/s, which corresponds to a Fourier frequency of the signal comb around 500 Hz. Two out-of-phase intensity interferograms are constructed at the two outputs of the FTS, with total power of 31.5% of the input power (16.5% in one output and 15% in the other). One of the output beams is directed to a cantilever-enhanced photoacoustic cell, while the other is directly measured with a HgCdTe detector (VIGO System) and used for normalization of the PA signal. For a typical input power of 50 mW, 8 mW out of the FTS is guided to the PA cell, of which 4.8 mW and 3.2 mW are contained in the signal and idler combs, respectively.

The photoacoustic cell (Gasera, PA201) with a cantilever-enhanced detector is used. The output beam of the FTS has a diameter of 5 mm and the PA cell has an input inner diameter of 4 mm. Therefore, a focusing lens ($f$ = 50 cm) was placed 25 cm in front of the PA cell to reduce the beam diameter to 2.5 mm. The PA cell is made of gold-coated aluminium with a length of 100 mm and a sample volume of 8 mL. A double-path configuration is implemented using a gold-coated mirror on the back side of the cell resulting in power spectral density of the signal comb of 42 μW/$cm^{-1}$. The displacement of the cantilever as a result of pressure changes due to absorption is measured via an integrated spatial-type interferometer in the photoacoustic cell.

Two mass flow controllers (Bronkhorst, F-201CV) and a pressure regulator (Bronkhorst, P-702CV) are used to flow the gas sample and regulate the pressure in the line leading to the PA cell. The PA cell has an internal pumping and gas exchange system that draws the sample from the supply line using micro pumps and automatic valve system. This gas supply system allows for changing the pressure inside the PA cell in the range between 300 mbar – 1000 mbar. All measurements were performed at 23 °C, as controlled by the PA cell thermostat. The test gas methane (100 ± 10 ppm in $N_2$, Air Liquide) was used. Pure $N_2$ was available for flushing the cell and diluting the methane sample.

To obtain broadband OFC-PAS spectra, three interferograms were acquired simultaneously using a National Instruments hardware (PCI-6221, 250 kS/s, 16 bit) and home-written LabVIEW™ program: the PA and the comb interferograms, as well as the cw laser interferogram for OPD calibration. Afterwards, the PA and the comb interferograms are resampled at the zero-crossings and extrema of the cw laser interferogram using home-written MATLAB® program. The absolute value of the Fourier transform of the OPD-calibrated interferograms yields the frequency-calibrated PA spectrum and the comb intensity envelope. Finally, the PA spectrum is divided by the comb intensity envelope to yield a normalized PA spectrum. The spectra of the C-H stretch band of $CH_4$ were measured at different working pressures of 1000 mbar, 650 mbar, and 400 mbar. At each pressure, the nominal resolution of the FTS was adjusted to yield 3 points per full width at half maximum (FWHM) of the methane lines, which resulted in a resolution of 1000 MHz, 650 MHz, and 400 MHz for the three pressures, respectively. The acquisition times of single interferograms with these resolutions were 200 s, 308 s, and 500 s, respectively.

## 3 Results and discussion

### Photoacoustic signal and noise considerations

The PA signal is produced by the non-radiative collisional relaxation in periodically excited molecules, producing periodic local heating at the modulation frequency, which can be detected by means of microphones. The PA signal, $S(\nu)_{PA}$, measured in volts, is related to absorption coefficient, $\alpha(\nu)$ (in $cm^{-1}$), as:

$$S(\nu)_{PA} = \alpha(\nu)\times\phi\left(S_m C_{cell} P(\nu)\eta\right) \quad (1)$$

where the scaling factor, $\phi$ (in $V/cm^{-1}$), is a function of the cantilever sensitivity, $S_m$ (in V/Pa), the PA cell response constant, $C_{cell}$ (in $Pa/cm^{-1}/W$), the exciting light power, $P(\nu)$ (in W), and the efficiency of the conversion of light energy into heat, $\eta$. Since the power of the comb source is a function of frequency, $\nu$, it should be monitored and used to normalize the PA signal. The absorption coefficient for $CH_4$ transitions is related to the temperature-dependent line strength, $S(T)$ [in $cm^{-1}$/(molecule $cm^{-2}$)], gas density, $[C]$ (in molecule/$cm^3$) and the normalized line shape function, $g(\nu)$ (in 1/$cm^{-1}$) as $\alpha(v) = S[C]g(v)$, where the line parameters can be found in the HITRAN database.[24]

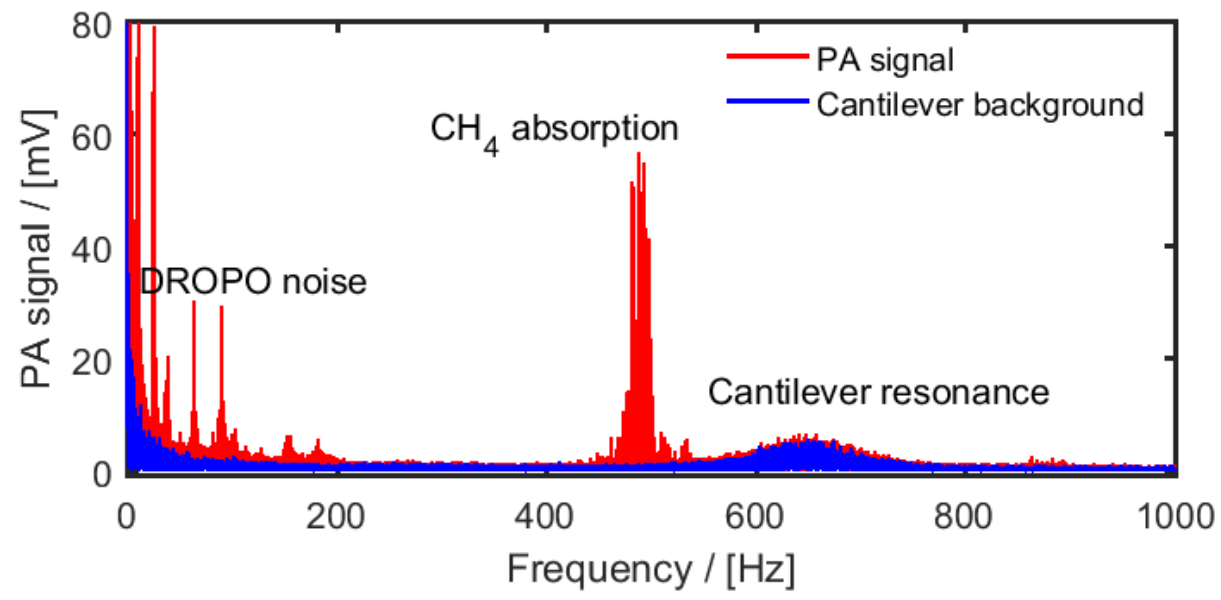

Fig. 2 The OFC-PAS spectrum in the radio/Fourier frequency domain. Red: the PA signal of 100 ppm $CH_4$ in $N_2$ (cell pressure: 1000 mbar, resolution: 1000 MHz). Blue: cantilever background measured with no light to the PA cell.

The intensity of the measured PA signal at a given total pressure changes linearly with gas density under the assumption that (i) the density of the absorbing gas is relatively low, so that Beer-Lambert law is linear, (ii) the input laser power is not high, so that absorption saturation effects can be neglected, and (iii) the relaxation rate from absorbed energy to heat is much faster than the rate of heating/cooling of the gas sample. However, the measured signal is also proportional to the scaling factor, $\phi$, which is a function of the sample pressure, composition, and temperature. Therefore, the measurement conditions of pressure and gas matrix have to be optimized to accomplish the highest $S/N$ according to the application requirements (vide infra).

Fig. 2 presents a typical OFC-PAS spectrum in the Fourier frequency domain. The red trace was recorded for 100 ppm $CH_4$ in $N_2$ at a total cell pressure of 1000 mbar and a resolution of 1000 MHz, while the blue trace is the background noise of the cantilever measured with no light into the PA cell. The $CH_4$ absorption appears in the Fourier frequency domain at ~500 Hz, which is determined by the OPD scan velocity. The broad peak at ~650 Hz is the background noise amplified by the cantilever resonance. The cantilever resonance amplifies both the signal and the noise levels equally, leaving the resulting $S/N$ unchanged.[13–15] The peaks at frequencies below 200 Hz are the intensity noise of the DROPO, which remains after stabilization of the DROPO cavity length. Therefore, the working range of the FTS modulation frequency that can be utilized for PAS measurements with constant cantilever noise level and high $S/N$ is 200 – 600 Hz. Note that this allows also the measurement of absorption at the idler comb frequencies, which would appear at ~300 Hz. Further improvements of the locking stability could minimize the DROPO intensity noise, extending the working range below the 200 Hz frequency.

**Proof-of-concept measurement**

Fig. 3 shows the OFC-PAS measurement of the C-H stretch band of 100 ppm of $CH_4$ in $N_2$ at 1000 mbar total pressure and 1000 MHz resolution. Panel a presents the normalized PA spectrum (red, left axis) of the *P*, *R*, and *Q* rovibrational structures of the band together with simulated absorption coefficient, $\alpha$ (blue, right axis), based on the parameters from the HITRAN database,[24] indicating the broadband coverage of the signal comb. A scaling factor $\phi$ = 10.5 V/cm$^{-1}$ was used in the simulations (see Eq. 1). Panel b is a zoom in around the Q-branch region, showing the high spectral resolution capability with no ILS distortion. Panel c shows the spectral envelope of the comb at 1000 MHz resolution (black), and a smoothing of the spectral envelope with 100 times lower resolution (10 GHz). The latter, interpolated to 1000 MHz, was used to normalize the acquired PA spectrum. Panels a and b show a very good agreement in the relative line intensities between the measured PA spectrum and the simulated absorption spectrum based on the HITRAN database.[24] The slight offset of the baseline in panels a and b is caused by taking the absolute value of the FFT.

We have observed that the entire C-H stretch band is imaged with a much smaller amplitude at the lower and higher wavelength sides symmetrically around the band origin. Such "ghost" imaging might be attributed to modulation of the intensity of the light source or sampling errors of the PA interferogram. These artefacts account for the slight intensity mismatch between the measurements and the simulations in the lower wavelength side and the unassigned lines in the higher wavelength side. It should also be noted that the normalization of the PA signal by the comb envelope may introduce a slight intensity mismatch with the simulations due to the absorption of molecular species (mainly interfering water lines) in the non-common beam path.

**Resolution and limit of detection**

Absorption features under the available pressure range of the PA cell (hundreds of millibars) have FWHM of hundreds of MHz. Thus, the nominal resolution of our FTS is sufficient to record the pressure broadened absorption features with negligible ILS contribution. Fig. 4 shows (red, left axis) the PA spectra of the C-H stretch band around the Q-branch region for 100 ppm $CH_4$ in $N_2$ at different total cell pressures and spectral resolutions: panel a - 1000 mbar and 1000 MHz resolution; panel b - 650 mbar and 650 MHz resolution; panel c - 400 mbar and 400 MHz resolution. A very good agreement between the experiment and the simulated absorption coefficient, $\alpha$ (blue, right axis), has been obtained at all working pressures. Scaling factors ($\phi$ in Eq. 1) were 10.5 V/cm$^{-1}$, 9.20 V/cm$^{-1}$, and 15.0 V/cm$^{-1}$ in the simulations for 1000 mbar, 650 mbar, and 400 mbar, respectively. Since the three measurements shown in Fig. 4 were performed with the same input laser power, PA cell, and gas matrix (i.e., *P*, $C_{cell}$ and $\eta$ are constants), the change of the scaling factor is a direct measure

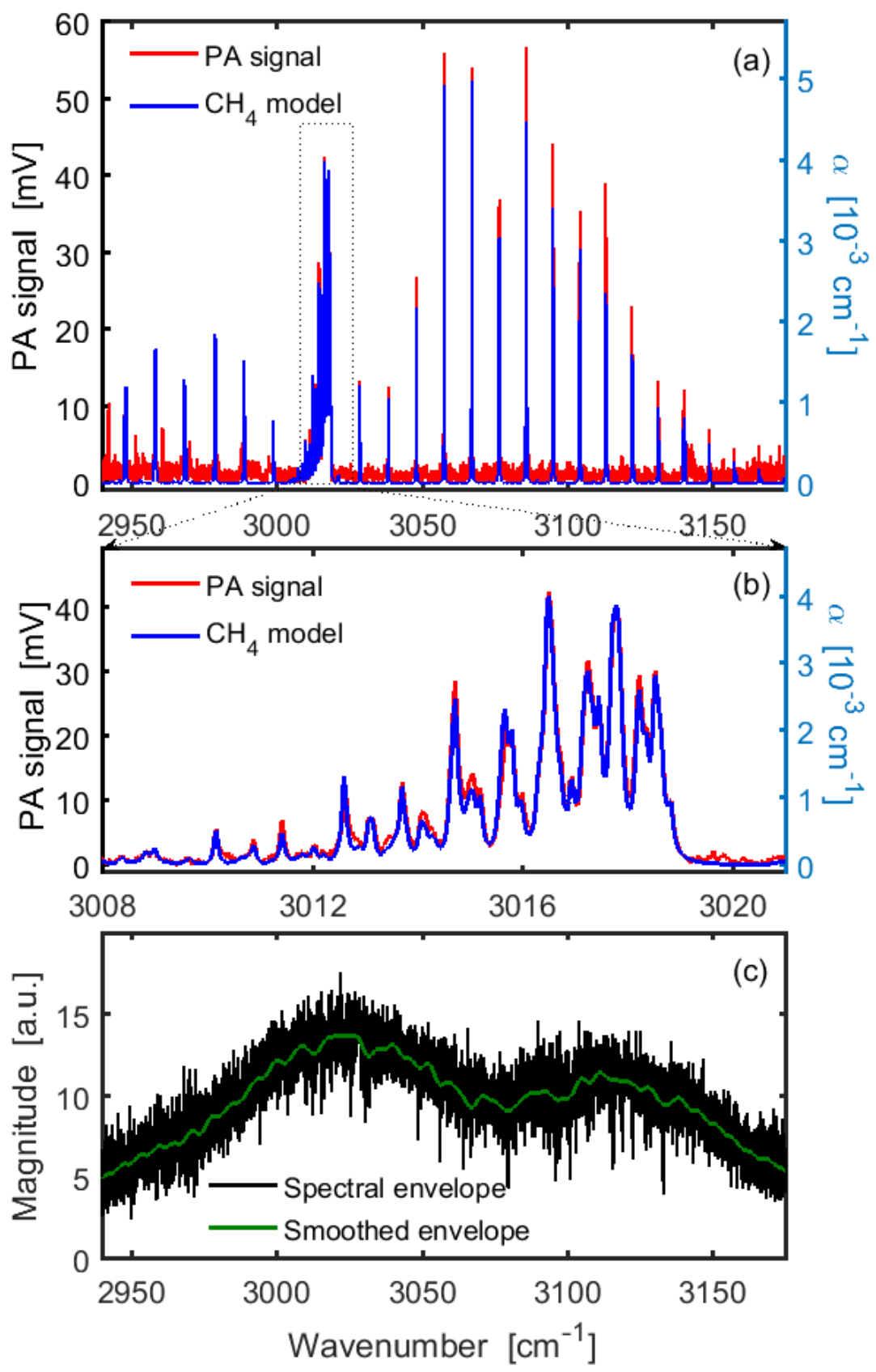

Fig. 3 (a) PA rovibrational spectrum of the C-H stretch band of 100 ppm of methane (red, left axis: 1000 mbar cell pressure, 1000 MHz resolution, 200 s measurement time) compared to the simulated absorption coefficient, $\alpha$, based on the HITRAN database (blue, right axis), (b) zoom of the Q-branch region, and (c) the spectral envelope of the comb.

of the change of cantilever sensitivity with pressure, attributed to the so-called gas spring and the effective mass that depend on the pressure and the molecular mass of the gas.

Table 1 summarizes the scaling factors, signal and noise levels as well as the *S/N* observed at the three different pressures. The noise, $\sigma(CH_4)$, is estimated as the standard deviation of the baseline of the $CH_4$ spectra around 3021 cm$^{-1}$. We note that the noise in the presence of $CH_4$ absorption is higher than the noise level measured with pure $N_2$ in the PA cell ($\sigma(N_2)$), also listed in Table 1). This difference might originate from the small structure that exists on the baseline of the $CH_4$ measurements due to the aforementioned "ghost" lines. We also note that background noise measurements with light on and off yielded similar levels. The *S/N* is calculated as the ratio of the line-centre PA signal around 3016.5 cm$^{-1}$ ($S_{PA}$) and the standard deviation of the noise on the baseline around 3021 cm$^{-1}$.

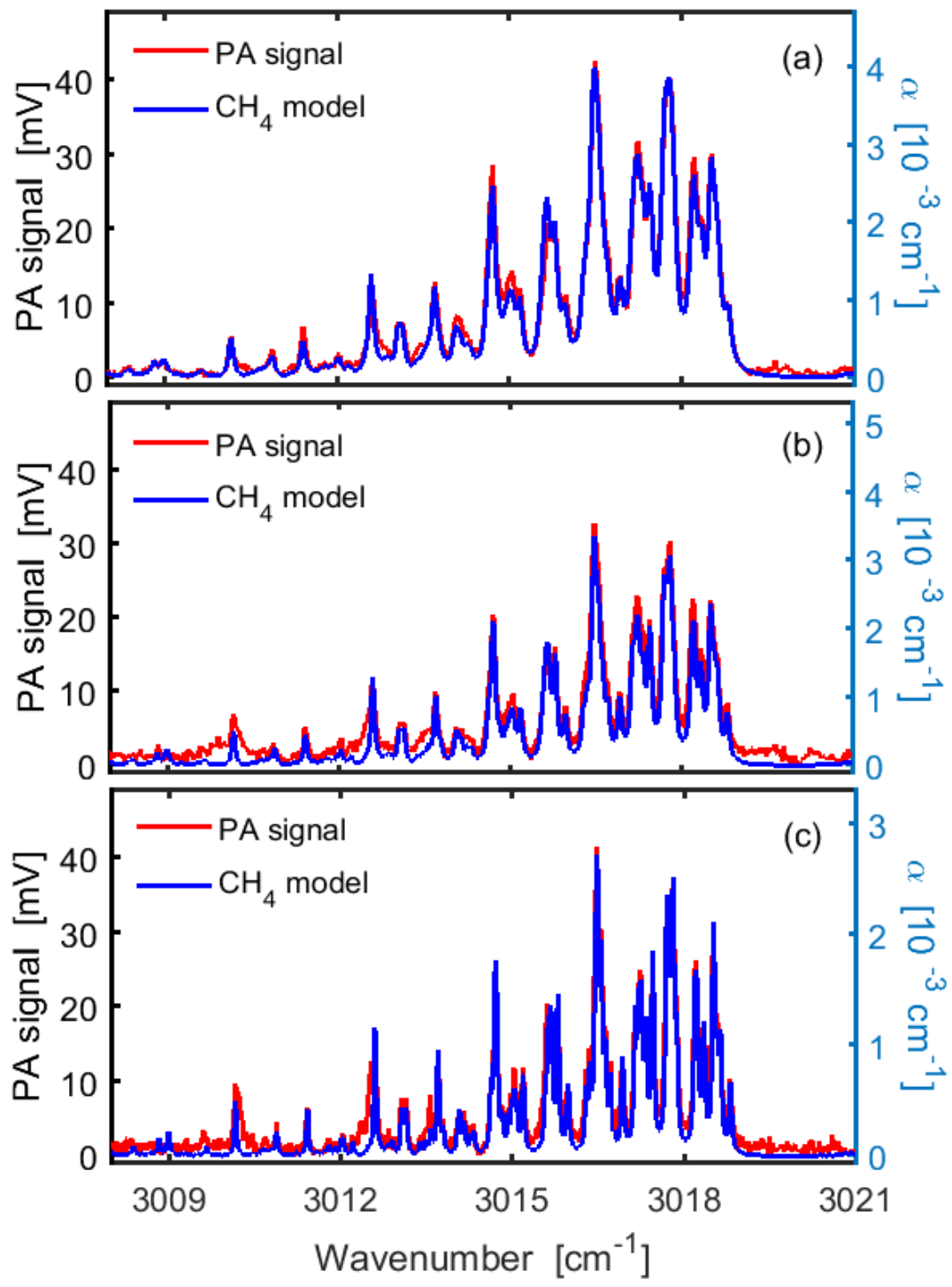

Fig. 4 PA spectra of the Q-branch of the C-H stretch band of 100 ppm methane (red, left axis) at (a) 1000 mbar and 1000 MHz resolution, (b) 650 mbar and 650 MHz resolution, and (c) 400 mbar and 400 MHz resolution, compared to models based on the HITRAN database (blue, right axis). Scaling factors of 10.5 V/cm$^{-1}$, 9.20 V/cm$^{-1}$, and 15.0 V/cm$^{-1}$ were used in the simulations for panels a, b, and c, respectively.

In conventional FT-IR the *S/N* decreases with the square root of the increase of the resolution,[26] assuming constant signal and white noise. The change of *S*/N must also account for the change of the line-centre absorption coefficient with pressure and number density. The line-centre absorption, also listed in Table 1, denoted by $F_\alpha$ and normalized to 1 at 1000 mbar, is estimated for the line at 3016.5 cm$^{-1}$ using the HITRAN database.[24] The last column of Table 1 shows the *S/N* expected in conventional FT-IR, taking into account the change of resolution and the line-centre absorption coefficient (assuming *S/N* of 100 at 1000 mbar). The discrepancy between the measured *S/N* and that predicted for conventional FT-IR is attributed to the fact that our PA measurements are not purely white noise limited, and that the cantilever signal and noise are enhanced by different factors when the pressure changes.

In general one can conclude that the fundamental trading rule of measurement time, resolution, and sensitivity of conventional FT-IR must be modified to include also the cantilever response and applied according to the application need. For example, high-resolution spectral measurements, which are essential for better selectivity and multicomponent analysis, would preferably be performed at the pressure of 400 mbar rather than 650 mbar, since there is an indication of higher cantilever sensitivity that compensates the decrease in *S/N* due to increased resolution and lower $F_\alpha$.

We evaluated the attainable limit of detection (LOD) based on the *S/N* of the strongest methane line at 3058 cm$^{-1}$ (see Fig. 3) measured at 1000 mbar, equal to 120. This corresponds to a detection limit of 0.8 ppm in 200 s for an exciting signal comb power spectral density of only 42 μW/cm$^{-1}$ in the absorption band of methane. Table 2 compares our LOD with that of previous FT-IR-PAS and the recent SC-PAS experiments using cantilever-enhanced detectors (all normalized to 1σ and 100 s measurement time). As shown in Table 2, the LOD is comparable for the three methods, while the resolution of OFC-PAS is more than two orders of magnitude better. Moreover, considering the lower power spectral density and the higher resolution of OFC-PAS compared to SC-PAS,[19] the attainable LOD of OFC-PAS becomes about a factor of five better than that of SC-PAS.

We also evaluated the performance of the system in terms of the normalized noise equivalent absorption, $NNEA = \alpha_{\min} P_{\text{sp.el.}} \sqrt{t}$, where $\alpha_{\min}$ is the minimum detectable absorption, equal to 4.2 × 10$^{-5}$ cm$^{-1}$ at 1000 mbar, $P_{\text{sp.el.}}$ is the power per single spectral component, and $t$ is the measurement time. At a resolution of 1000 MHz (0.033 cm$^{-1}$), the power spectral density of 42 μW/cm$^{-1}$ corresponds to $P_{\text{sp.el.}}$ of 1.4 μW. Considering the measurement time of 200 s, we obtained a NNEA of 8 × 10$^{-10}$ W cm$^{-1}$ Hz$^{-1/2}$, which is a factor of two better than typical values obtained for PAS with cw lasers,[27] and a factor of three worse than the best reported value using cw lasers.[7] The very low NNEA obtained here confirms that the comb source is not causing significant noise contribution. Moreover, the capability of OFC-PAS to record spectrum over thousands of elements simultaneously is not fully reflected in the definition of NNEA.

Table. 1 Scaling factors, experimental signal and noise levels as well as the *S/N* at different pressures and spectral resolutions, compared to S/N expected for conventional FT-IR.

| Pressure [mbar] | Resolution [MHz] | $\phi$ [V/cm$^{-1}$] | $S_{PA}$ [a] [mV] | σ($CH_4$) [mV] | σ($N_2$) [mV] | Measured S/N | $F_\alpha$ [b] | Expected S/N |
|---|---|---|---|---|---|---|---|---|
| 1000 | 1000 | 10.5 | 42.8 | 0.43 | 0.23 | 100 | 1.0 | 100 |
| 650 | 650 | 9.20 | 31.6 | 0.61 | 0.32 | 52 | 0.85 | 69 |
| 400 | 400 | 15.0 | 41.2 | 0.74 | 0.52 | 56 | 0.69 | 44 |

[a] line-centre PA signal at 3016.5 cm$^{-1}$
[b] normalized line-centre absorption coefficient

Table. 2 The obtained limits of detection, LOD (in ppm, for 1σ and 100 s), spectral resolution, and power spectral density, $\tilde{P}$, of our experiment compared to that of FT-IR-PAS and the recent SC-PAS using cantilever-enhanced detector.

| | LOD [ppm] | Res. [GHz] | $\tilde{P}$ [μW/cm$^{-1}$] | Ref. |
|---|---|---|---|---|
| FT-IR-PAS | 0.50 | 120 | | 18 |
| FT-IR-PAS | 1.9 | 240 | | 15 |
| SC-PAS | 0.33 | 120 | 61 | 19 |
| OFC-PAS | 1.1 | 1 | 42 | This work |

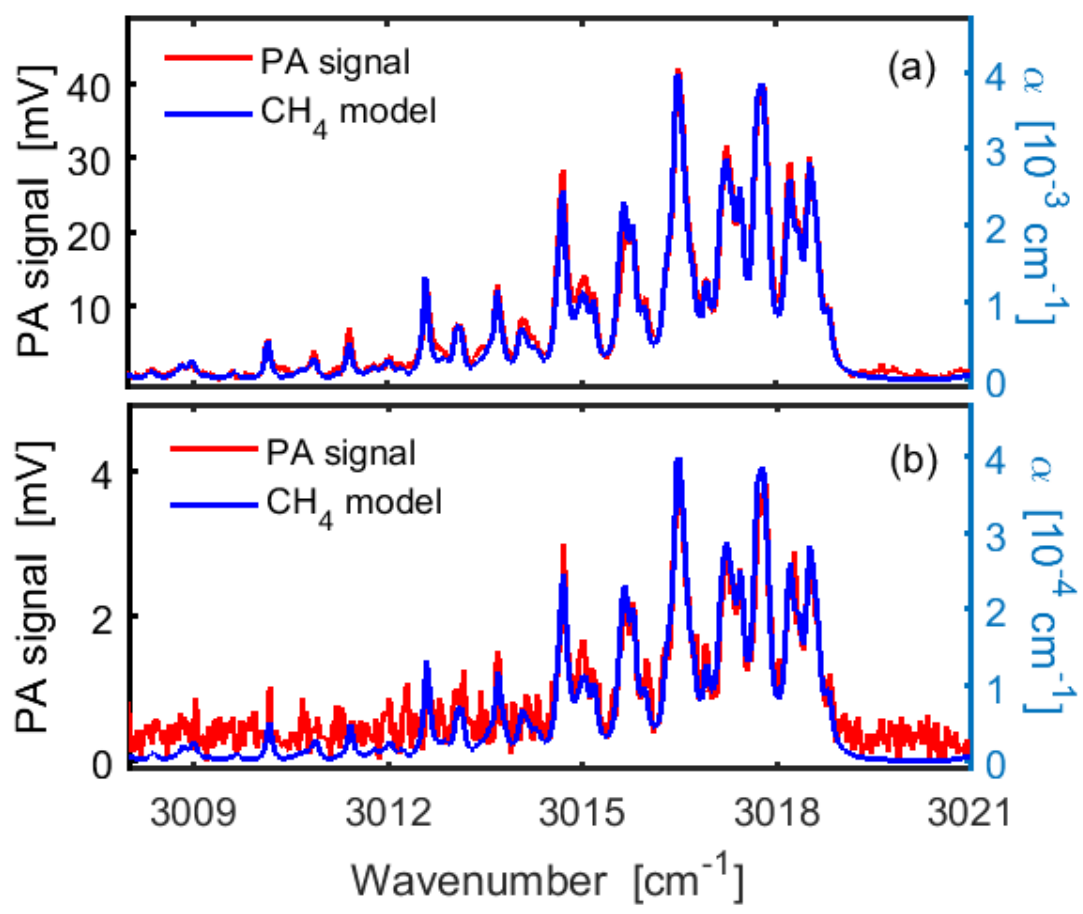

Fig. 5 PA spectrum of the C-H stretch band of methane around the Q-branch region at 1000 MHz resolution for 100 ppm methane in $N_2$ (panel a, left axis) and 10 ppm methane in $N_2$ (panel b, left axis). Simulations (right axis) are based on HITRAN database using a scaling factor of 10.5 V/cm$^{-1}$.

**Linearity and dynamic range**

The linearity of the recorded PA signal for different number density at the same total cell pressure is demonstrated in Fig. 5, which shows spectra of 100 ppm (panel a) and 10 ppm methane in $N_2$ (panel b) measured at the same total pressure of 1000 mbar and resolution of 1000 MHz. The same scaling factor of 10.5 V/cm$^{-1}$ has been used in both simulations (blue, right axis), giving a very good agreement with the experiment (red, left axis) in both cases. We also measured the PA spectrum for mixing ratio of 1% $CH_4$ in $N_2$ (data is not shown). No saturation effects on the measured PA interferogram have been observed (i.e., the centre burst of the interferogram was not cut), indicating a four orders of magnitude dynamic range (i.e. from $10^{-6}$ to $10^{-2}$). However, at this high mixing ratio a non-linearity of the signal has been observed due to non-linearity of the Beer-Lambert law. Therefore, it can be concluded that the presented OFC-PAS is linear over two orders of magnitude within a dynamic range of four orders of magnitude.

## 4 Conclusions

The broadband optical frequency comb photoacoustic spectroscopy (OFC-PAS) technique has been demonstrated for the first time. A Fourier trasnform spectrometer is used to modulate the comb intensity and a cantilever-enhanced detector is used in a non-resonant mode to record the photacoustic signal over a wide frequency range in a small sample volume.The PA spectrum is normalized by the comb intensity envelope that is measured simultaneously at the second output of the FTS. In the first demonstration of OFC-PAS, we measured the PA spectra of the fundamental C-H stretch band of $CH_4$ in $N_2$ at total pressures between 400 – 1000 mbar, with no instrumental line shape distortion and in a very good agreement with simulated absorption spectra based on the HITRAN database. While the obtained detection limit of methane of 0.8 ppm in 200 s, with power spectral density of 42 μW/cm$^{-1}$, is comparable to that achieved in previous broadband PAS demonstrations based on incoherent and supercontinuum light sources, the resolution is two orders of magnitude better, limited only by the pressure broadening of the methane absorption lines. Moreover, we obtained a normalized noise equivalent absorption of 8 × $10^{-10}$ W cm$^{-1}$ Hz$^{-1/2}$, which is only a factor of three higher than the best reported with continuous wave laser PAS.

In the measurements presented in the current work, the nominal resolution of the FTS was higher than the repetition rate of the comb source, $f_{rep}$, and sufficient to record the absorption lines with negligible ILS distortion. If needed, spectral features narrower than $f_{rep}$ can be measured with no ILS distortion using the method of comb-based FTS with sub-nominal resolution, in which interferograms with length matched precisely to $c/f_{rep}$ are measured.[23,28,29]

The detection limit of OFC-PAS can be improved by several measures. First of all, a factor of 40 improvement can be expected using available high-power mid-IR OFC sources[30] with an output power of up to 1.5 W, compared to 36 mW of the signal comb used in this work. Secondly, using multi-line fitting routines instead of the single point *S/N* determination would decrease the uncertainty of the fitted concentration and yield improved LOD, where a factor of 11 improvement is expected for the C-H stretch band of methane addressed in our work.[31] Using these two measures combined, a LOD of methane in the lower ppb range should be readily accomplished.

The high fidelity of the high-resolution broadband spectra measured with OFC-PAS holds the potential to extend the capability of optical spectrometers for multispecies trace gas analysis in very small sample volumes.

## Acknowledgements

The authors thank Amir Khodabakhsh and Chuang Lu for help in optimizing the DROPO performance, Isak Silander for help in setting up the gas supply system, and Francisco Senna Vieira, Teemu Tomberg and Juho Karhu for useful discussions about OFC-PAS. The work at UmU is financed by the Knut and Alice Wallenberg Foundation (KAW 2015.0159). M.V. and T.M. acknowledge the financial support of the Academy of Finland (Grant 314363).

## References

1 F. J. M. Harren, J. Mandon and S. M. Cristescu, *Photoacoustic Spectroscopy in Trace Gas Monitoring: Encyclopedia of Analytical Chemistry*, John Wiley and Sons: Chichester, U.K., 2012.

2 T. Schmid, *Anal. Bioanal. Chem.*, 2006, **384**, 1071–1086.

3 A. G. Bell, *Am. J. Sci.*, 1880, **20**, 305–324.

4 E. L. Kerr and J. G. Atwood, *Appl. Opt.*, 1968, **7**, 915–921.

5 J. Li, W. Chen and B. Yu, *Appl. Spectrosc. Rev.*, 2011, **46**, 440–471.

6 V. Spagnolo, P. Patimisco, S. Borri, G. Scamarcio, B. E. Bernacki and J. Kriesel, *Opt. Lett.*, 2012, **37**, 4461–4463.

7 T. Tomberg, M. Vainio, T. Hieta and L. Halonen, *Sci. Rep.*, 2018, **8**, 239–245.

8 Y. He, Y. Ma, Y. Tong, X. Yu and F. K. Tittel, *Opt. Express*, 2018, **26**, 9666–9675.

9 R. E. Lindley., A. M. Parkes, K. A. Keen, E. D. McNaghten and A. J. Orr-Ewing, *Appl. Phys. B*, 2007, **86**, 707–713.

10 A. A. Kosterev, Y. A. Bakhirkin, R. F. Curl and F. K. Tittel, *Opt. Lett.*, 2002, **27**, 1902–1904.

11 K. Liu, X. Guo, H. Yi, W. Chen, W. Zhang and X. Gao, *Opt. Lett.*, 2009, **34**, 1594–1596.

12 P. Patimisco, G. Scamarcio, F. K. Tittel and V. Spagnolo, *Sensors*, 2014, **14**, 6165–6206.

13 K. Wilcken and J. Kauppinen, *Appl. Spectrosc.*, 2003, **57**, 1087–1092.

14 T. Kuusela and J. Kauppinen, *Appl. Spectrosc. Rev.*, 2007, **42**, 443–474.

15 J. Uotila and J. Kauppinen, *Appl. Spectrosc.*, 2008, **62**, 655–660.

16 V. Koskinen, J. Fonsen, K. Roth and J. Kauppinen, *Vib. Spectrosc.*, 2008, **48**, 16–21.

17 K. H. Michaelian, *Photoacoustic IR spectroscopy. Instrumentation, applications and data analysis*, Wiley-VCH, Weinheim, 2nd edn., 2010.

18 C. B. Hirschmann, J. Uotila, S. Ojala, J. Tenhunen and R. L. Keiski, *Appl. Spectrosc.*, 2010, **64**, 293–297.

19 T. Mikkonen, C. Amiot, A. Aalto, K. Patokoski, G. Genty and J. Toivonen, arXiv:1807.00895.

20 P. R. Griffiths and J. A. de Haseth, *Fourier transform infrared spectrometry*, Wiley-Interscience; Chichester : John Wiley [distributor], Hoboken, N.J., 2nd edn., 2007.

21 V. Werwein, J. Brunzendorf, A. Serdyukov, O. Werhahn and V. Ebert, *J. Mol. Spectrosc.*, 2016, **323**, 28–42.

22 J. Mandon, G. Guelachvili and N. Picqué, *Nat. Photonics*, 2009, **3**, 99 -102.

23 L. Rutkowski, A. C. Johansson, G. Zhao, T. Hausmaninger, A. Khodabakhsh, O. Axner and A. Foltynowicz, *Opt. Express*, 2017, **25**, 21711–21718.

24 L. Rothman, I. Gordon, Y. Babikov, A. Barbe, D. C. Benner, P. Bernath, M. Birk, L. Bizzocchi, V. Boudon, L. Brown, A. Campargue, K. Chance, E. Cohen, L. Coudert, V. Devi, B. Drouin, A. Fayt, J.-M. Flaud, R. Gamache, J. Harrison, J.-M. Hartmann, C. Hill, J. Hodges, D. Jacquemart, A. Jolly, J. Lamouroux, R. L. Roy, G. Li, D. Long, O. Lyulin, C. Mackie, S. Massie, S. Mikhailenko, H. Mu ¨ller, O. Naumenko, A. Nikitin, J. Orphal, V. Perevalov, A. Perrin, E. Polovtseva, C. Richard, M. Smith, E. Starikova, K. Sung, S. Tashkun, J. Tennyson, G. Toon, V. Tyuterev and G. Wagner, *J. Quant. Spectrosc. Radiat. Transfer*, 2013, **130**, 4–50.

25 A. Khodabakhsh, V. Ramaiah-Badarla, L. Rutkowski, A. C. Johansson, K. F. Lee, J. Jiang, C. Mohr, M. E. Fermann and A. Foltynowicz, *Opt. Lett.*, 2016, **41**, 2541–2544.

26 N. R. Newbury, I. Coddington and W. Swann, *Opt. Express*, 2010, **18**, 7929–7945.

27 J. Peltola, M. Vainio, T. Hieta, J. Uotila, S. Sinisalo, M. Metsälä, M. Siltanen and L. Halonen, *Opt. Express*, 2013, **21**, 10240–10250.

28 P. Maslowski, K. F. Lee, A. C. Johansson, A. Khodabakhsh, G. Kowzan, L. Rutkowski, A. A. Mills, C. Mohr, J. Jiang, M. E. Fermann and A. Foltynowicz, *Phys. Rev. A*, 2016, **93**, 021802/1-5.

29 L. Rutkowski, P. Masłowski, A. C. Johansson, A. Khodabakhsh and A. Foltynowicz, *J. Quant. Spectrosc. Radiat. Transfer*, 2018, **204**, 63–73.

30 F. Adler, K. C. Cossel, M. J. Thorpe, J. Hartl, M. E. Fermann and J. Ye, *Opt. Lett.*, 2009, **34**, 1330–1332.

31 F. Adler, P. Masłowski, A. Foltynowicz, K. C. Cossel, T. C. Briles, I. Hartl and J. Ye, *Opt. Express*, 2010, **18**, 21861–21872.